\documentclass[a4paper,11pt]{article}
\usepackage{jheppub} 
\usepackage{lineno}

\usepackage{pbox}
\usepackage{color}
\usepackage{soul}

\title{\boldmath Existence of Vacuum Wormholes in Einsteinian Cubic Gravity }

\author[a]{Mengqi Lu}
\author[a,b]{Jiayue Yang}
\author[a,b]{Robert B. Mann}

\affiliation[a]{University of Waterloo, \\
200 University Ave W, Waterloo, Canada}
\affiliation[b]{Perimeter Institute for Theoretical Physics, \\ 31 Caroline ST. N., Waterloo, Canada}

\emailAdd{m64lu@uwaterloo.ca}\emailAdd{jiayue.yang@uwaterloo.ca} \emailAdd{rbmann@uwaterloo.ca}

\abstract{
Wormhole solutions in gravitational theories typically require exotic matter.
Here we present a wormhole solution to
the field equations of Einsteinian Cubic Gravity
--- a phenomenological competitor to general relativity that includes terms cubic in the curvature --- 
that has no matter, exotic or otherwise. These purely gravitational wormhole geometries 
are asymptotically AdS but contain a geometric deficit at infinity. The deficit, interpreted as a global monopole, plays an essential role in our construction. We find that our wormhole solution satisfies traversablility criteria. We also find, for different parameters, a range of possible wormhole solutions. }

\begin{document}
\maketitle
\flushbottom
\section{Introduction}
\label{sec:intro}
Wormholes, a fundamental concept in the study of spacetime travel, have been widely discussed since their formal introduction within the framework of General Relativity (GR) in 1935 \cite{Einstein:1935tc}. However, ensuring the traversability of wormholes within GR necessitates the presence of exotic matter \cite{Morris:1988cz}. This requirement has spurred research into modified theories of gravity \cite{DeFalco:2023twb,Harko:2013yb,Fierro:2018rna,Dehghani:2011vu,Myrzakulov:2015kda,Battista:2024gud,DeFalco:2021ksd,DiGrezia:2017daq}, as such theories often incorporate additional structures that may enable the construction of traversable wormholes without the need for exotic matter \cite{Ellis:1973yv, Ellis:1979bh, Bronnikov:1973fh,Rosa:2022osy,Rosa:2023guo}. 

Our particular interest is in Generalized Quasi-Topological Gravity (GQTG) \cite{Hennigar:2017ego, Bueno:2019ycr, Bueno:2022res, Moreno:2023rfl,Hennigar:2016gkm,Bueno:2016lrh,Bueno:2016xff,Hennigar:2018hza}
as they form a class of interesting generalizations of general relativity.
These theories are characterized by
their  field equations 
yielding  a single metric function (i.e. $g_{tt}g_{rr}=-1$) for a static spherically symmetric ansatz, and are ghost-free on constant-curvature backgrounds. The   higher-curvature terms introduce additional degrees of freedom, which enriches the thermodynamic properties of black holes \cite{Bueno:2017qce,Hennigar:2017umz,Mir:2019rik,Mir:2019ecg,Lu:2023hgu}. Most importantly, unlike Lovelock theories \cite{Lovelock:1970zsf,Lovelock:1971yv} or quasi-topological gravities \cite{Oliva:2010eb,Myers:2010ru,Oliva:2012zs,Oliva:2011xu,Dehghani:2011vu,Cisterna:2017umf}, GQTGs have non-trivial field equations
in four dimensions \cite{Hennigar:2017ego, Bueno:2019ycr, Bueno:2022res, Moreno:2023rfl}. Einsteinian Cubic Gravity (ECG) \cite{Bueno:2016xff,Hennigar:2017ego}, as the simplest non-trivial theory in GQTGs, has an  action of the general form 
\begin{equation}\label{action}
    \mathcal{I}=\frac{1}{16\pi}\int d^d x \sqrt{-g}(-2\Lambda_0+ R+\alpha \mathcal{P}+\beta \mathcal{C} +\gamma \mathcal{C}'),
\end{equation} 
where $\alpha$, $\beta$ and $\gamma$ are the respective couplings of the curvature densities $\mathcal{P}$, $\mathcal{C}$ and $\mathcal{C}'$, with $\Lambda_0$ the cosmological constant. These cubic corrections are given by 
 \begin{equation}
    \mathcal{P}=12R_{a\ b}^{\ c\ d}R_{c\ d}^{\ e\ f }R_{e\ f}^{\ a\ b}+R_{ab}^{\ \ cd}R_{cd}^{\ \ ef}R_{ef}^{\ \ ab}-12R_{abcd}R^{ac}R^{bd}+8{R^b}_{a}{R^c}_{b}{R^a}_{c},
\end{equation}
\begin{equation}
    \mathcal{C}=\frac{1}{2}R{R_a}^{b}{R_b}^{a}-2R^{ac}R^{bd}R_{abcd}-\frac{1}{4}RR_{abcd}R^{abcd}+R^{de}R_{abcd}R^{abc}_{\ \ \ \  e}.
\end{equation}
\begin{equation}
    \mathcal{C}'={R_a}^b{R_b}^c{R_c}^a-\dfrac{3}{4}R{R_a}^b{R_b}^a+\dfrac{1}{8}R^3.
\end{equation}

Our primary result   is the discovery of a numerical wormhole solution described by a general static spherically symmetric ansatz  \begin{equation}\label{GSSS}
ds^2=g_{tt}(r)dt^2+\frac{dr^2}{g^{rr}(r)}+r^2(d\theta^2+\sin^2{\theta}d\phi^2), 
\end{equation}
within the framework of 4D ECG in the absence of any matter. This implies that the overall effect of gravity must become repulsive near the wormhole throat;  otherwise a timelike geodesic congruence would converge and forbid the formation of a throat \cite{Poisson:2009pwt}.  Indeed, we find 
(see Appendix~\ref{sec:Construction}) that 
the size of the throat marginally increases with
increasing mass parameter, indicative of
a repulsive effect.
The solution we found connects two copies of an asymptotically AdS spacetime with a spherical deficit/surfeit angle $\delta \in (-1,0)$,
\begin{equation}\label{bcatlarge}
 r\rightarrow \infty: \quad  g_{tt}\sim - \bigg(\frac{r^2}{l^2}+1+\delta\bigg)+\mathcal{O}(r^{-1}), \quad g^{rr}\sim \frac{r^2}{l^2}+1+\delta+\mathcal{O}(r^{-1}),
\end{equation}
analogous to that of a global monopole \cite{Barriola:1989hx}. This choice of the asymptotic form validates the local analysis of series solutions as it distinguishes the formal series solutions at infinity of a wormhole from its AdS black hole counterparts \cite{Lu:2024bdw}, which turns out to be crucial in the construction. 

Before proceeding further, we note for the
ansatz \eqref{GSSS} that
$\mathcal{C}$ and $\mathcal{C}'$ are linearly dependent. Henceforth we can set
$\gamma=0$  without loss of generality.


\section{Basic Strategy}\label{sec:BS}

Our construction relies  on the shooting method, namely, looking for initial conditions satisfying \eqref{bcatlarge} at $r=\infty$ such that the wormhole boundary conditions  are fulfilled at the throat $r_{\mathrm{th}}$. For this purpose, we take $u\equiv 1/r,\  F(u)\equiv -1/g_{tt}, \ n(u)\equiv -g^{rr}/g_{tt}$ to rewrite the metric ansatz \eqref{GSSS} as 
\begin{equation}\label{Newansatz}
    ds^2=-\frac{1}{F(u)}dt^2+\frac{1}{u^4}\frac{F(u)}{n(u)}du^2+\frac{1}{u^2} (d\theta^2+\sin^2{\theta}d\phi^2).
\end{equation}
Then the metric functions satisfying \eqref{bcatlarge} are regularized at the initial point $u=0$ ($r=\infty$), as the initial conditions \eqref{bcatlarge} become
\begin{equation}\label{newbcatlarge}
  u\rightarrow 0: \quad  F\sim l^2u^2-(1+\delta)l^4u^4+\mathcal{O}(u^5),\quad n\sim 1+ \mathcal{O}(u^3).
\end{equation}

The boundary conditions for a wormhole throat $u_{\mathrm{th}}\equiv 1/r_{\mathrm{th}}$ involve more subtleties. First, the Morris-Throne condition \cite{Morris:1988cz} is equivalent to $n=0$   and $F$ being positive finite at $u=u_{\mathrm{th}}$. Second, since the throat should be a regular point in the spacetime and the non-trivial Riemannian tensor components in a tetrad depend on $n',F',F''$ only, it suffices to require these derivatives to be finite there. 

However, these boundary conditions are more troublesome than they appear to be. The two third order field equations (given in
\eqref{feqno} in Appendix~\ref{ad:feqns}),  obtained from varying the gravitational action \eqref{action} and using  the ansatz \eqref{Newansatz}, are   linear in $F^{(3)}$ and $n^{(3)}$. These equations can thus be decomposed into $6$ first-order equations
\begin{equation}\label{feqns}
    p'=\xi(F,n,s,p,v,w),\  w'=\zeta(F,n,s,p,v,w),\  s\equiv F', \ p\equiv s', \ v\equiv n', \ w \equiv v',
\end{equation}
where $\xi$ and $\zeta$ are algebraic functions defined as
\begin{equation}
    \begin{aligned}
      & \mathcal{A} \xi\equiv 8 F^7 (u^2-\Lambda_0)-6 F^2 n u^8 \{s^2 u^2 v^2 (58 \alpha +7 \beta )+8 \alpha  n^2 (p^2 u^2+26 p s u+54 s^2)\\&+4 n s u [4 p u v (9 \alpha +\beta )+24 \alpha  s^2+s
   u w (10 \alpha +\beta )+2 s v (62 \alpha +5 \beta )]\}\\&+3 F^3 u^7 \{4 n^2 u
   [ 48s \alpha  p u+2su w (16 \alpha +\beta )+4s v (40 \alpha +\beta )+pu^2 w (8 \alpha +\beta )\\&+8pu v (10 \alpha +\beta
   )+132 \alpha  s^2]+128 \alpha  n^3 (2 p u+3 s)-s u^3 v^3 (8 \alpha +\beta )\\&+2 n u^2 v [2 p u v (8 \alpha +\beta )+48 \alpha  s^2+3 s u w (8
   \alpha +\beta )+2 s v (44 \alpha +5 \beta )]\}\\&+6 F^4 u^7 \{4 n^2 (-32 \alpha  p
   u-48 \alpha  s+\beta  u w+4 \beta  v)+4 \alpha  s u^2 v^2\\&+n u [\beta  v (8 s+v)-16 \alpha  (p u v+s u w+5 s v)]\}\\&+2 F^5
   u^3 \{3 \beta  u^5 v^2-4 n [s+3 \beta  u^4 (u w+4 v)]\}\\&+120 F n^2 s^2 u^9 [4 \alpha  n (p u+4 s)+s u v (10 \alpha +\beta )]-8 F^6 n u^2-480 \alpha  n^3 s^4 u^{10},
   \\
   \\&\mathcal{B}\zeta\equiv 6 F^2 u^5 \{-4 \alpha  n^2 [10 v (p u+5 s)+u w (p u+16 s)]-s u^2 v^3 (9 \alpha +\beta )\\&-2 n u v [p u v (9 \alpha +\beta )+24 \alpha  s^2+s u w (21
   \alpha +2 \beta )+6 s v (14 \alpha +\beta )]\}\\&+3 F^3 u^4 \{16 \alpha  n^2 (7 u
   w+8 v)+2 n u [4 \alpha v (6 p u+30 s+31 v)+4 \alpha u w (6 s+25 v)\\&+8 \alpha u^2 w^2+\beta  (u^2 w^2+10 u v w+7
   v^2)]+u^2 v^2 [24 \alpha  s+u w (8 \alpha +\beta )+2 v (10 \alpha +\beta )]\}\\&-12 F^4 u^4 [4 \alpha  n (7 u
   w+8 v)+u v (2 \alpha  u w+4 \alpha  v-\beta  v)]\\&+30 F n s u^6 [4 \alpha  n (p u v+s u w+8 s v)+s u v^2 (12
   \alpha +\beta )]-240 \alpha  n^2 s^3 u^7 v-4 F^5 v,
    \end{aligned}
\end{equation}
with denominators $\mathcal{A}$ and $\mathcal{B}$ of $\xi$ and $\zeta$, respectively given by
\begin{equation}
    \mathcal{A}\equiv 12 F^2 n^2 u^9 \left[8 \alpha  F^2- 8 \alpha  nF-u v F (8 \alpha +\beta )+4 \alpha  n s u\right], \quad \mathcal{B}\equiv \mathcal{A}/(2u^3n).
\end{equation}
From the expressions for $\mathcal{A}$ and $\mathcal{B}$, the Morris-Throne condition 
$n(u_{\mathrm{th}})=0$  
implies that  $\xi$ and $\zeta$ simultaneously diverge. A wormhole throat is therefore generically a spontaneous singularity for the system \eqref{feqns}, 
which is why we must ensure that the derivatives $n',F',F''$ 
are finite there, in turn ensuring the  curvature is finite. In practice, the most annoying issue is that a singularity yields large numerical errors in the solution in its vicinity, especially for higher derivatives, which forbids us for obtaining those derivatives directly through numerical solutions, such as shooting methods usually do. We thus have to introduce another approach. For now, let us  take a look at the initial value problems that we have to deal with.

\section{Initial Value Problems}

In general, shooting methods require tuning $6$ parameters for our system. This number is halved if we limit our consideration to analytic local solutions at $u=0$, namely, metric functions having the form 
\begin{equation}\label{seriesatu0}
    F(u)=F_2u^2+F_4u^4+\sum_{i=5}^{
    \infty} F_i u^i, \quad n(u)=1+\sum_{i=3}^{\infty}n_i u^i,
\end{equation}
at $u=0$. Solving
 \eqref{feqns} order by order in $u$ in a neighbourhood of $0$, we find 
\begin{equation}\label{satinfy}
    \Lambda_0 + 3/F_2 + 24\alpha /F_2^3 = 0, \quad \Lambda_0 + 2/F_2 = 0,
\end{equation}
from the lowest two orders \cite{Lu:2024bdw}. Note that these equations imply that there is no
$\alpha\to 0$ limit for the solutions we obtain.

The coefficients $F_{i\geq 6}$ and $n_{i\geq 3}$ can be expressed in terms of $F_2,F_4,F_5, \beta$; a few of these coefficients  are explicitly presented in   Appendix \ref{ad:expatu0}. The physical interpretations of $F_2,F_4, F_5$ are given by 
\begin{equation}\label{meaning}
    F_2\equiv l^2>0, \quad  F_4\equiv -(1+\delta)l^4\in (-l^4,0), \quad F_5\equiv \frac{4l^4(2l^4-9\beta)}{9(l^4-2\beta)}M,
\end{equation}
where $F_2$ and $F_4$ are determined such that the boundary condition \eqref{newbcatlarge} is satisfied; $F_5$ is proportional to the ADM mass $M$ of the spacetime, which is identified as negative one-half the coefficient of $1/r$ in $g^{rr}$. Therefore, since $\Lambda_0$ and $\alpha$ can be expressed in  terms of $F_2$ via \eqref{satinfy}, the $6$ values of metric functions and their lowest two derivatives are specified once  $F_2,F_4,F_5 $ is fixed.

We emphasize that the above dependence of $F_{i\geq 6}$ and $n_{i\geq 3}$ on $F_2,F_4, F_5$ is only valid when $1/F_2 \neq 0$ and $\delta \neq 0$. If either condition is not satisfied, the analyticity \eqref{seriesatu0} of both metric functions at $u=0$ forces every $n_{i\geq3}$ to be zero \cite{Lu:2024bdw}, which implies $n$ is a constant function in the common convergent region of the series for $F$ and $n$, and further indicates the product of $g_{tt}g_{rr}$ is a constant there. 
 Arbitrary choices of $F,F',F''$ at $u=0$ lead to  black hole solutions when $g_{tt}g_{rr}$ is constant for the whole range of $r$. This implies, 
 for initial conditions that are the same as for a given black hole solution, 
 that wormholes can't exist when $g_{tt}g_{rr}$ is a constant in any neighbourhood of $u=0$. 
  This makes us particularly interested in asymptotically AdS solutions with non-zero $\delta$, since
this condition is required to obtain  wormholes.

In practice, we introduce a tiny deviation parameter $\epsilon > 0$ to numerically integrate from $u=\epsilon$ to larger $u$ with initial values $F(\epsilon),n(\epsilon),F'(\epsilon),n'(\epsilon),F''(\epsilon),n''(\epsilon)$ instead. The point of this is to avoid the trouble caused by the  generic singularity at $u=0$ leading to $\mathcal{A}=\mathcal{B}=0$. In our construction $\epsilon$ is taken to be $0.00056$. The part of the solution with $u<0.00056$ can be obtained by integrating from $0.00056$ to $0$. The functions 
$n(u)$ and $F(u)/u^2$ both
approach constants
(matching \eqref{seriesatu0} perfectly) until  they eventually terminate at some rather tiny value limited by precision. 

\section{Wormhole Identification}\label{sec:Identification}

\begin{figure}
    \centering
     \includegraphics[width=9.cm]{  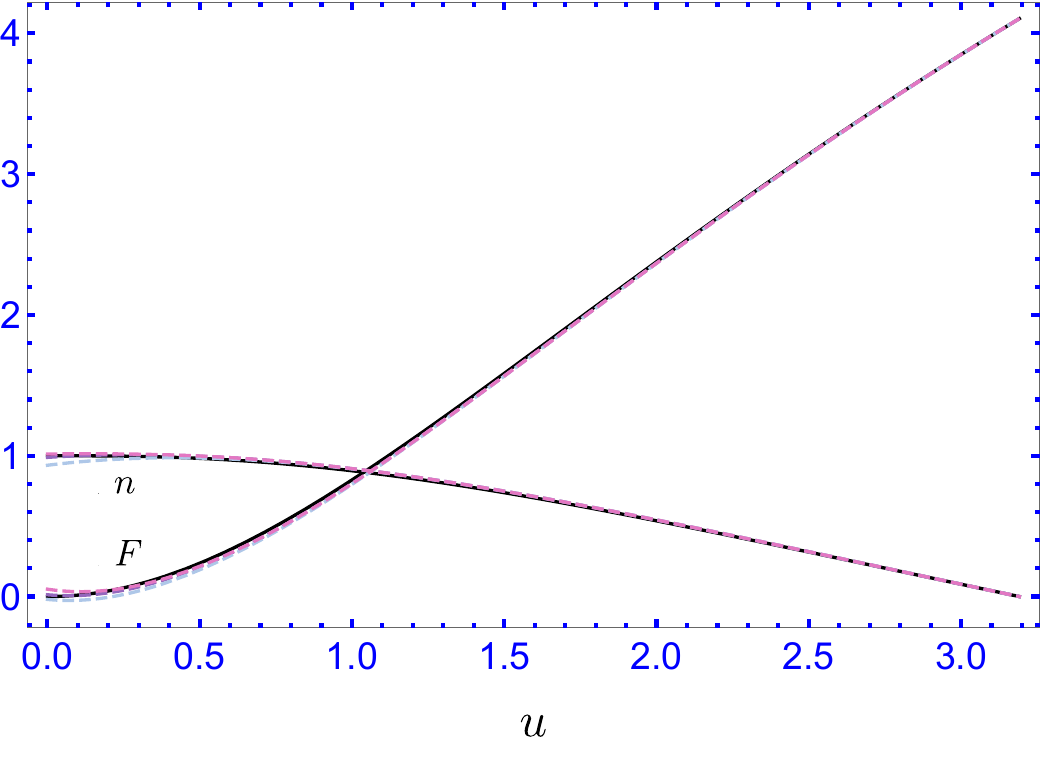}  
    \caption{A wormhole solution with a positive mass $M$ \eqref{meaning} is constructed subject to $\beta=0.091, F_2=1, F_4\approx-0.281186 \in (-1,0), F_5=0.1, u_{\mathrm{th}}\approx 3.19114$. The increasing  and decreasing solid black curves correspond to the numerical solutions for $F(u)$ and $n(u)$ respectively. Dashed curves are truncations  of the series 
    \eqref{seriesatth}. Truncation orders are $7$, $12$, $20$ for the respective blue, pink, and purple curves for $F$; and $8$, $13$, $21$ for  $n$. We see that that pink and purple series curves  overlap with the numeric ones for almost the entire domain of $u$, indicating the speed of convergence for the series \eqref{ssol}.}
    \label{fig:wormhole}
\end{figure}

Figure \ref{fig:wormhole} illustrates a numerical wormhole solution (solid curves), obtained   from solving the initial value problem at $u=0$ for a typical choice of $F_2,F_4,F_5 $ and $\beta$. The wormhole throat $u_{\mathrm{th}}$ is located at the place where $n$ vanishes. Both metric functions terminate at $u=u_{\mathrm{th}}$, since $u_{\mathrm{th}}$ is generically a singular point, which causes numerical errors in evaluating higher derivatives of metric functions there. 
Consequently we need an alternative means for identifying    $u_{\mathrm{th}}$ as a wormhole throat.  

To this end, we first check the possibility that a series solution local at $u_{\mathrm{th}}$ not only satisfies desired boundary conditions discussed in Section \ref{sec:BS} but also converges to the numerical solution at the terminal point. The local solution discovered in \cite{Lu:2024bdw} is a Taylor series of the form 
\begin{equation}\label{seriesatth}
F(u)=\sum_{i=0}^{\infty}\tilde{F}_i (u-u_{\mathrm{th}})^i, \quad n(u)=\sum_{i=1}^{\infty}\tilde{n}_i (u-u_{\mathrm{th}})^i,
\end{equation} 
with coefficients obtained by solving the equations \eqref{feqns} order by order in $(u-u_{\mathrm{th}})$. Similar to \eqref{seriesatu0}, higher order coefficients are uniquely characterized by the lower order ones $\{ \tilde{F}_0, \tilde{n}_1, F_2,\beta, u_{\mathrm{th}}\}$, as indicated in Appendix \ref{ad:expatth}. Thus $5$ parameters in total need to be specified to obtain the series solution.

The dashed curves in Figure \ref{fig:wormhole} represent different truncated series of the local solution \eqref{seriesatth} at $u_{\mathrm{th}}=3.19114$, which reads
\begin{equation}\label{ssol}
\begin{aligned}
    &F=4.10434 +1.30746(u-u_{\mathrm{th}})-0.133104 (u-u_{\mathrm{th}})^2 0.00189318 (u-u_{\mathrm{th}})^3\\&+0.00389799(u-u_{\mathrm{th}})^4 -0.00140705(u-u_{\mathrm{th}})^5 +0.000345987(u-u_{\mathrm{th}})^6 +\cdots,
    \\&n=-0.456446 (u-u_{\mathrm{th}})+0.0130613(u-u_{\mathrm{th}})^2 +0.00718277   (u-u_{\mathrm{th}})^3\\&-0.00229803(u-u_{\mathrm{th}})^4 +0.000442028 (u-u_{\mathrm{th}})^5 -0.0000556576(u-u_{\mathrm{th}})^6 +\cdots.
\end{aligned}
\end{equation}
The values of $F_2,\beta, u_{\mathrm{th}}$ are taken from
the numeric solution (shown
as  the solid curve), whereas the values of $\tilde{F}_0, \tilde{n}_1$ are induced from $F(u_{\mathrm{th}})$ and $n'(u_{\mathrm{th}})$ from the numerical solution. 
We find that even
if we truncate the series
at relatively low orders
(blue dashed curves in 
Figure \ref{fig:wormhole}) 
the convergence with the
numerical solution
is very good, even near $u=0$. 

The close match between the series solution \eqref{ssol} and the numeric solution near the throat gives us confidence that 
the solid curve
is indeeed a numeric wormhole solution to the field equations. Improved accuracy can be obtained
by further tuning the initial conditions to
obtain a better match between the series 
and numeric solutions.

\section{Traversability of the Wormhole}

Since the series \eqref{ssol} provides a very good approximation of higher order derivatives near throat, we can use it  to investigate the traversability \cite{Radhakrishnan:2024rnm,Morris:1988cz} of the numerical solution described in Figure \ref{fig:wormhole}. 

Due to the power series form of \eqref{ssol}, the factor $(F/n)^{1/2}$ of the proper radial distance $l(u)\equiv \int_{u_{\mathrm{th}}}^{u} du (F/n)^{1/2}/u^2$ is no longer ill-defined at $u_{\mathrm{th}}$. This can be seen by considering a small perturbation $\Delta u<0$ around the throat, which implies $ l \approx \sqrt{F(u_{\mathrm{th}})\Delta u/n'(u_{\mathrm{th}})}/u^2_{\mathrm{th}}$ being regular. Therefore $l(u)$ is finite for all $0<u \leq u_{\mathrm{th}}$. 

The Riemannian curvature tensor at the throat in an orthonormal basis has only four independent non-vanishing components 
\begin{equation}
\begin{aligned}
&R_{\hat{t}\hat{r}\hat{t}\hat{r}}=\frac{4nF'^2-n'FF'-2nFF''}{4F^3} =0.918443,\quad R_{\hat{\theta}\hat{\phi}\hat{\theta}\hat{\phi}} = \frac{F-n}{Fr^2}=10.1834\\&R_{\hat{r}\hat{\theta}\hat{r}\hat{\theta}}=R_{\hat{r}\hat{\phi}\hat{r}\hat{\phi}}=-\frac{1}{2r}\bigg(\frac{n}{F}\bigg)'=-1.80698, \quad R_{\hat{t}\hat{\theta}\hat{t}\hat{\theta}}=R_{\hat{t}\hat{\phi}\hat{t}\hat{\phi}}=-\frac{nF'}{2F^2r}=0,
\end{aligned}
\end{equation}
where $'$ indicates derivatives with respect to $r$. All of them are finite, implying regularity of the spacetime and of the tidal force experienced by a traveler moving at a slow speed when passing through the throat \cite{Morris:1988cz}.

The flare-out condition \cite{Morris:1988cz}
is easily visualized by  considering the embedding of the wormhole geometry for a fixed $t$ and $\theta=\pi/2$ into an AdS space 
\begin{equation}
    ds^2=\bigg(1+\delta + \frac{\rho^2}{l^2}\bigg)dz^2+\bigg(1+\delta + \frac{\rho^2}{l^2}\bigg)^{-1}d\rho^2+\rho^2 d\psi^2,
\end{equation}
with a deficit angle $\delta$ 
through the embedding relation 
\begin{equation}
    \rho=r, \quad \psi=\phi, \quad dz=\frac{\sqrt{F/n(1+\delta+r^2/l^2)-1}}{(1+\delta+r^2/l^2)} dr .
\end{equation}
This implies the throat truly corresponds to a local minimum in the radial  coordinate since
\begin{equation}
    \frac{d}{dz}\bigg(\frac{dr}{dz}\bigg)\bigg|_{u_{\mathrm{th}}}=1.02888>0
\end{equation}
satisfying the flare-out condition  $\frac{d}{dz}\frac{dr}{dz}|_{u_{\mathrm{th}}} > 0$.

\section{Conclusions and Future Work}

In   general relativity, traversable wormholes solutions typically require the presence of exotic matter to exist. Here we have obtained for the first time  a wormhole solution that requires no exotic matter  in the framework of generalized quasi-topological gravity. Specifically, in Figure \ref{fig:wormhole}, 
we have shown that
a traversable wormhole 
connecting two asymptotically AdS spacetimes 
exists as a solution
to the field equations of  4D Einsteinian Cubic
Gravity (ECG). The spacetime has  a geometrical deficit 
$\delta$, which can be
interpreted as a global monopole.  
We also present some wormhole candidates in  Appendix~\ref{sec:Construction}.  Note that the coupling $\alpha$ is chosen to be a specific value \cite{Lu:2024bdw} and coupling $\beta$ is non-zero in all the constructions.

During the course of our investigation, by varying the initial and boundary conditions, we also observed the possibility of numerical soliton solutions within the framework of 4D ECG.  The metric for these solutions  has neither an  horizon nor a throat; the functions $F(u)$ and  $n(u)$ both asymptotically approach finite values as $u\to\infty$. This new type of solution warrants further investigation; rather than present them  here, we leave this for future study. 

Another interesting aspect to consider is the exploration of 4D GQTG wormhole
solutions that are asymptotically flat or de Sitter (dS). These solutions are of greater physical interest.

\acknowledgments
This work was supported in part by the  Natural Science and Engineering Research Council of Canada. Mengqi Lu  was also supported by China Scholarship Council. Research at Perimeter Institute is supported in part by the Government of Canada through the Department of Innovation, Science and Economic Development and by the Province of Ontario through the Ministry of Colleges and Universities.
\appendix

\section{Wormhole Candidates}\label{sec:Construction}

\begin{figure}
    \centering
    \includegraphics[width=9cm]{  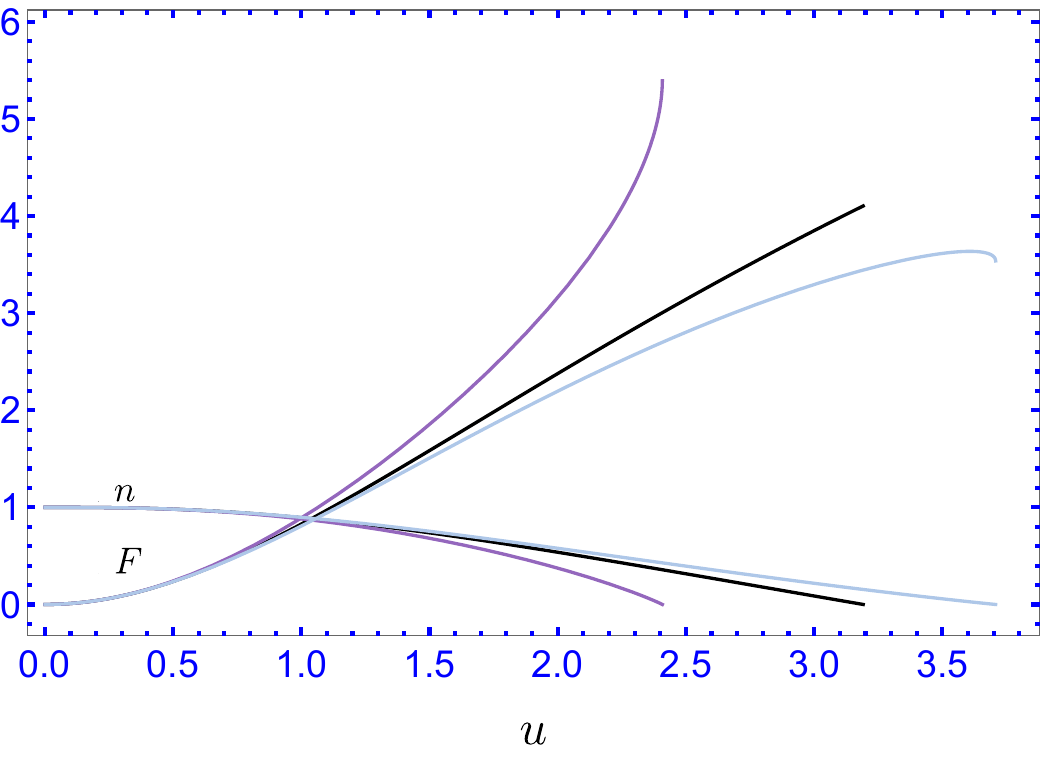}
    \caption{Solutions for different $F_4$'s. Raising and descending curves of same color respectively correspond to $F(u)$ and $n(u)$ of a solution. The increasing and decreasing curves originating at $u=0$  respectively correspond to $F(u)$ and $n(u)$.   The black is the wormhole solution in Figure \ref{fig:wormhole} with $\beta=0.091, F_2=1, F_4\approx-0.281186, F_5=0.1$. The purple and the blue only differ the black in $F_4$ (purple: $F_4=-0.1$; blue: $F_4=-0.31$).}
    \label{fig:variations}
\end{figure}

Obtaining the value of $F_4$ in Figure \ref{fig:wormhole} requires a trick in making the termination point of numerical methods ``less singular". Given $\beta, F_2$ and $F_5$, a generic value of $F_4$ causes a ``tip" in $F(u)$, bending it either up or down in the vicinity of a singularity, as shown in Figure \ref{fig:variations} in blue or purple. This ``tip" should be avoided as it is clearly not a part of a Taylor series 
expansion near the throat
and implies a huge value for $F''(u_{\mathrm{th}})$. The value of $F_4$ for the black curve in Figure \ref{fig:variations} is determined by demanding that $F''(\tilde{u}_{\mathrm{th}})$ has a small value at a point $\tilde{u}_{\mathrm{th}}$ close to $u_{\mathrm{th}}$. For simplicity, we take $F''(0.9999999 u_{\mathrm{th}})$ to be $0$. However other values less than $\sim 10^3$ will not make the result much different.

\begin{figure}
    \centering
    \includegraphics[width=6.5cm]{  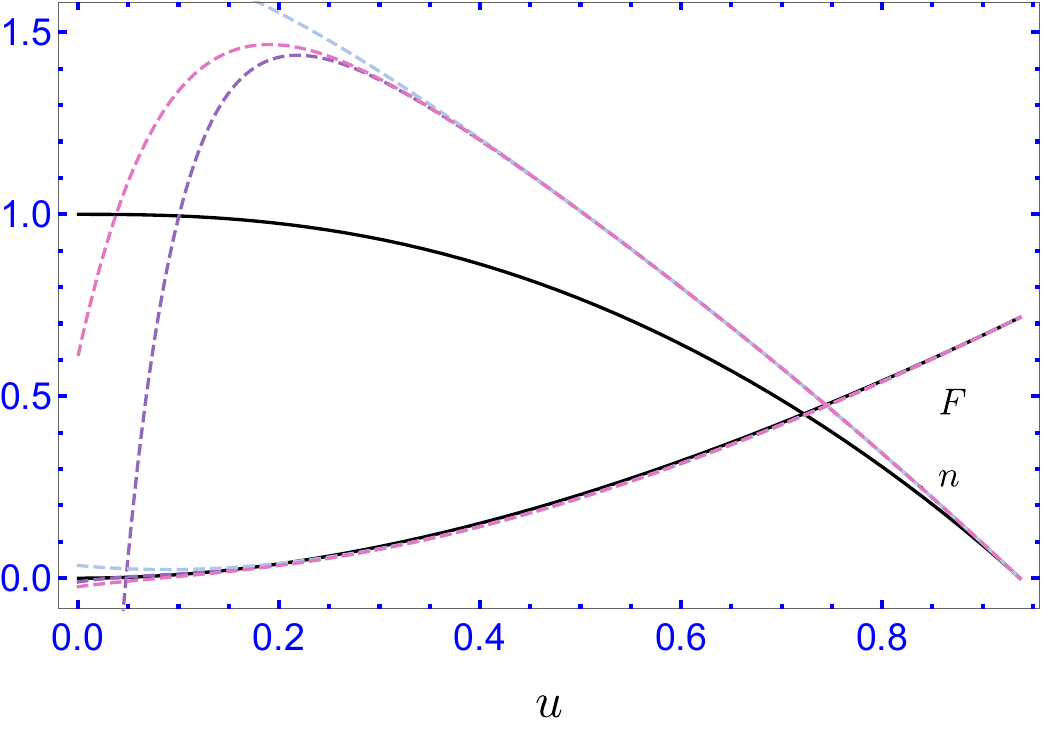}  \includegraphics[width=6.5cm]{  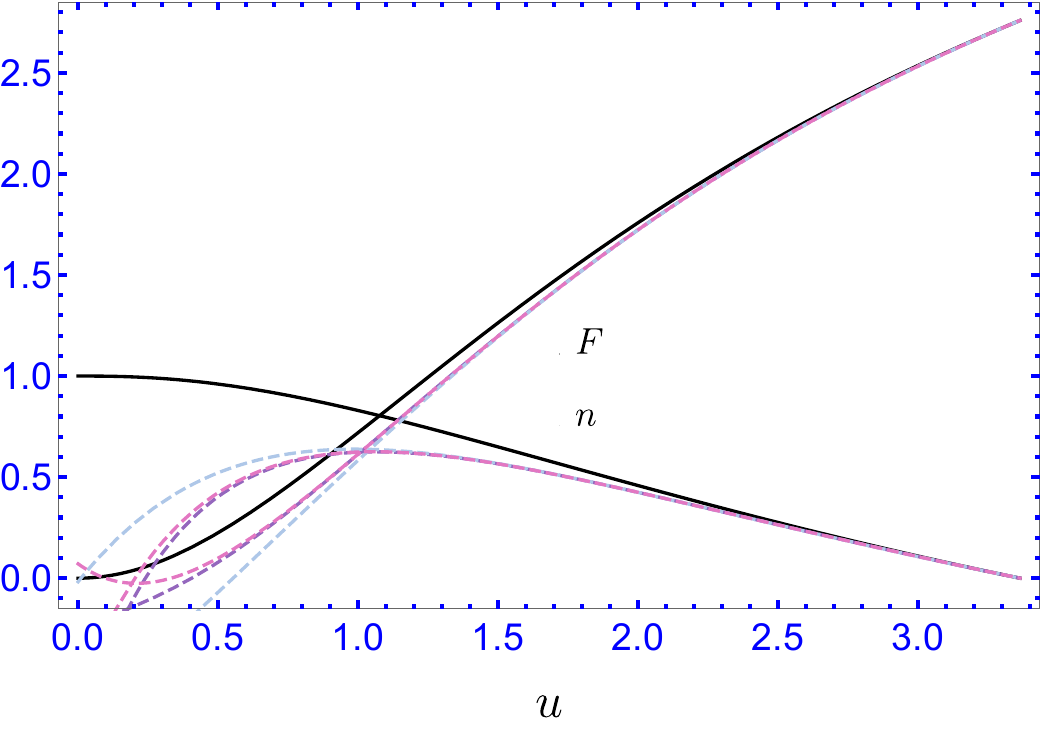} 
    \caption{Wormhole candidates. The same labeling convention as in Figure \ref{fig:wormhole} is applied here. Left: $\beta=0.2,F_2=F_5=1,F_4\approx-0.635778, u_{\mathrm{th}}\approx 0.937102$; Right: $\beta=0.05,F_2=F_5=1,F_4\approx-0.648749, u_{\mathrm{th}}\approx 3.36473$. }
    \label{fig:wormholes}
\end{figure}
\begin{figure}
    \centering
    \includegraphics[width=9cm]{  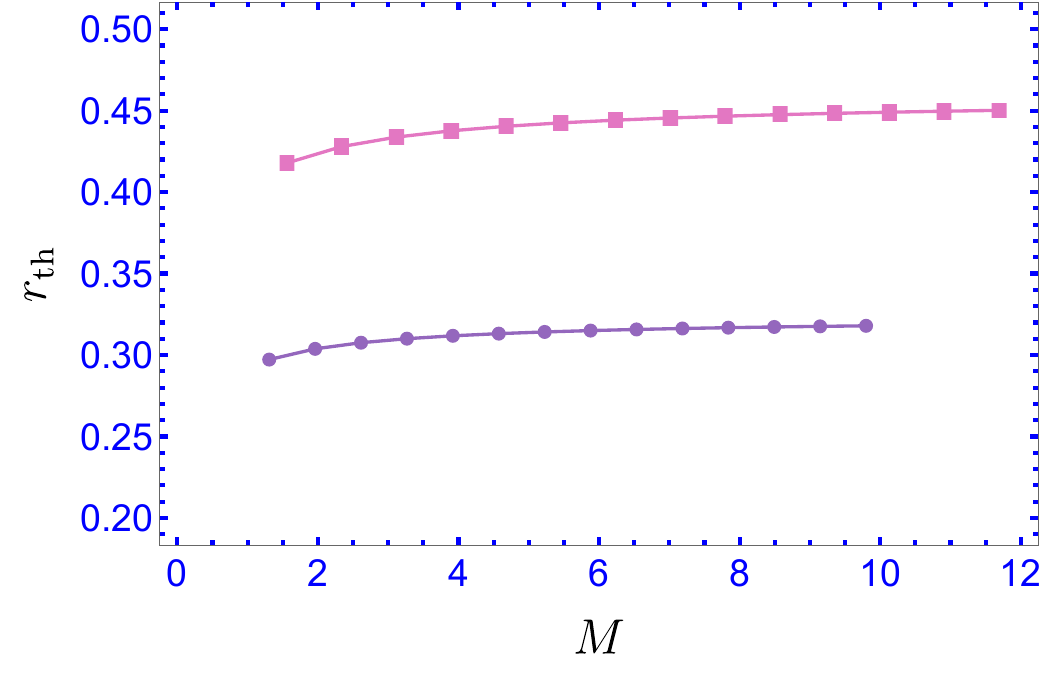} 
    \caption{Relations between wormhole mass $M$ and throat size $r_{\mathrm{th}}$ for $\beta=0.05$ (purple) and $0.091$ (pink). Both of them have the same $F_2=1$, and $F_4$ is obtained by the prescription in Section \ref{sec:Construction}.}
    \label{fig:rth_mass}
\end{figure}

   
Two wormhole candidates with differing values of   $\beta$ and $F_5$ 
from those in in Figure \ref{fig:wormhole} 
are shown in Figure \ref{fig:wormholes} based on the prescription for $F_4$ discussed above. From the figure, we see in each case  that there is
a segment of the  series solution that converges to the corresponding numerical solution (near the throat) but otherwise departs from it.
Since non-analytic solutions are allowed in  non-linear systems, a mismatch with the series \eqref{seriesatth} does not forbid the numeric solutions from being wormholes. Since $F'',F', n'$ appear to be finite from the figure, these solutions may also be regarded as  wormholes. If we relax the wormhole criterion and regard all solutions with a vanishing ``tip" to be wormholes, we arrive at Figure \ref{fig:rth_mass}, which displays the size of the throat as a function of the mass $M$. We see that the throat slowly
expands as $M$ increases, apparently asymptoting to an upper bound.  

\section{Field Equations}
\label{ad:feqns}

The equations of motion $\mathcal{E}_F =0 = \mathcal{E}_n$ for the two metric functions in \eqref{Newansatz} are obtained by the variation of the on shell action with respect to $F$ and $n$ respectively, where the explicit forms of $\mathcal{E}_F$ and $\mathcal{E}_n$ are given by
\begin{equation}\label{feqno}
\begin{aligned}
   &4F^7 \mathcal{E}_F/u^3\equiv 3 F^3
   u^4 \{2 n u [24 \alpha u n' F''+10 (10 \alpha +\beta ) u n'n''+u^2 n'n^{(3)}  (8 \alpha +\beta )\\&+24 \alpha  F'
   (u n''+5 n')+(124 \alpha +7 \beta ) n'^2+u^2 (8 \alpha +\beta ) n''^2]\\&+u^2
   n'^2 [24 \alpha  F'+2 (10 \alpha +\beta ) n'+u (8 \alpha +\beta ) n'']+16 \alpha  n^2 [u
   (n^{(3)} u+7 n'')+8 n']\}
    \\&+30 u^6 n F   F' \{u (12 \alpha +\beta ) F' n'^2+4 \alpha  n [u F'' n'+F' (u n''+8
   n')]\}-240 \alpha  n^2 u^7 F'^3 n'
   \\&-6 F^2 u^5 \{u^2 (9 \alpha +\beta ) F' n'^3+4 \alpha  n^2 [F' (n^{(3)}
   u^2+16 u n''+50 n')+u F'' (u n''+10 n')]\\&+2 n u n' [24 \alpha  F'^2+ 6 (14
   \alpha +\beta ) n'F'+u (21 \alpha +2 \beta ) F'n''+u (9 \alpha +\beta ) F'' n']\}\\&-12 F^4 u^4 \{u n'
   [(4 \alpha -\beta ) n'+2 \alpha  u n'']+4 \alpha  n [u (n^{(3)} u+7 n'')+8 n']\}-4 F^5 n',
   \\
   \\&8F^7n\mathcal{E}_{n}\equiv 8 F^7 (\Lambda_0-u^2) -120 F n^2 u^9 F'^2 [u (10 \alpha +\beta ) F' n'+4 \alpha  n (u F''+4 F')]\\&+6 F^2 n u^8
  \{u^2 (58 \alpha +7 \beta ) F'^2 n'^2+8 \alpha  n^2 [u^2 F''^2+u F'
   (F^{(3)} u+26 F'')+54 F'^2]\\&+4 n u F' [24 \alpha  F'^2+2 (62 \alpha
   +5 \beta )F' n'+u (10 \alpha +\beta )F' n''+4 u (9 \alpha +\beta ) F'' n']\}\\&+6 F^4 u^7 \{-4 \alpha  u^2 F'
   n'^2+4 n^2 [48 \alpha  F'+u (32 \alpha  F''+4 \alpha  F^{(3)} u-\beta  n'')-4 \beta 
   n']\\&+n u [ 8(10 \alpha -1 \beta ) F'n'+16 \alpha  u n'F''-\beta  n'^2+16 \alpha  u F' n'']\}\\&+F^5 \{8 n u^3 [F'+3 \beta
    u^4 (u n''+4 n')]-6 \beta  u^8 n'^2\}+8 F^6 n u^2+480 \alpha  n^3 u^{10} F'^4
    \\&-3
   F^3 u^7 \{2 n u^2 n' [48 \alpha  F'^2+ 2 (44 \alpha +5 \beta ) n'F'+3 u (8 \alpha +\beta )F'
   n''+2 u (8 \alpha +\beta ) F'' n']\\&+4 n^2 u [132 \alpha 
   F'^2+  48 \alpha  u F'F''+4 (40 \alpha +\beta ) F'n'+2u (16 \alpha +\beta ) F'n''\\&+  8
   (10 \alpha +\beta )u n' F''+u^2 (8 \alpha +\beta ) n''F''+F^{(3)} u^2 (8 \alpha +\beta ) n']-u^3 (8 \alpha +\beta ) F' n'^3\\&+32 \alpha  n^3 [u
   (F^{(3)} u+8 F'')+12 F']\}.
\end{aligned}
\end{equation}

\section{Taylor Series Solution at \texorpdfstring{$u=0$}{}}
\label{ad:expatu0}

The field equations imply that the series ansatz \eqref{seriesatu0} and \eqref{satinfy} have the
coefficients given in \eqref{seriesatu0}. These all depend only on $F_2,F_4,F_5$ and $\beta$, with a few of the leading terms being
\begin{equation}
    n_3=\frac{5F_2 F_5}{18\beta - 4 F_2^2}
\end{equation}
\begin{equation}
    n_4=\frac{135F_2 F_5^2 (3F_2^2-11\beta)}{16(9\beta -2F_2^2)^2(F_2^2+F_4)}
\end{equation}
\begin{equation}
    n_5=\frac{27 F_2 F_5^3 \left(11916 \beta ^2-6621 \beta  F_2^2+919 F_2^4\right)}{64 \left(F_2^2+F_4\right){}^2 \left(9 \beta -2
   F_2^2\right){}^3}+\frac{3 F_4 F_5}{9 \beta -2 F_2^2}
\end{equation}
\begin{equation}
\begin{aligned}
    n_6 &=\frac{45 F_2 F_5^4 \left(-3217212 \beta ^3+2700468 \beta ^2 F_2^2-755109 \beta  F_2^4+70339 F_2^6\right)}{448
   \left(F_2^2+F_4\right)^3 \left(9 \beta -2 F_2^2\right)^4}
   \\& \qquad \qquad \quad+  \frac{25 F_5^2 \left[F_2^2 \left(96 \beta +175 F_4\right)-630 \beta  F_4-22 F_2^4\right]}{112 \left(F_2^2+F_4\right) \left(9
   \beta -2 F_2^2\right)^2}
\end{aligned}
\end{equation}
\begin{equation*}
    \cdots
\end{equation*}
\begin{equation}
    F_6=\frac{F_4^2}{F_2}-\frac{9 F_5^2 \left(828 \beta ^2-363 \beta  F_2^2+37 F_2^4\right)}{32 \left(F_2^2+F_4\right) \left(9 \beta
   -2 F_2^2\right)^2}
\end{equation}

\begin{equation}
    F_7=\frac{9 F_5^3 \left(268164 \beta ^3-184383 \beta ^2 F_2^2+40020 \beta  F_2^4-2641 F_2^6\right)}{224
   \left(F_2^2+F_4\right){}^2 \left(9 \beta -2 F_2^2\right){}^3}+\frac{F_4 F_5 \left(34 F_2^2-135 \beta \right)}{7 \left(2
   F_2^3-9 \beta  F_2\right)}
\end{equation}
\begin{equation*}
    \cdots.
\end{equation*}

\section{Taylor Series Solution at \texorpdfstring{$u=u_{\mathrm{th}}$}{}}
\label{ad:expatth}

Similarly, we can obtain the coefficients in 
the series 
\eqref{seriesatth} by solving \eqref{feqno} order by order. A few coefficients are given by
\begin{equation}
    \begin{aligned}
        \tilde{F}_1=-\frac{1}{\tilde{n}_1^2 u_{\mathrm{th}}^9\mathcal{D}}[6 \tilde{F}_0^2 \beta  \tilde{n}_1^2 F_2 u_{\mathrm{th}}^8+8 \tilde{F}_0^4 \left(F_2 u_{\mathrm{th}}^2+2\right)]
    \end{aligned}
\end{equation}
\begin{equation}
    \begin{aligned}
        \tilde{n}_2&=\frac{1}{4 \tilde{n}_1 F_2 u_{\mathrm{th}}^9\mathcal{D}^2}[4 \tilde{F}_0^3 \tilde{n}_1 u_{\mathrm{th}} (48 \beta
   +24 \beta  F_2 u_{\mathrm{th}}^2-7 F_2^3 u_{\mathrm{th}}^2-18 F_2^2)\\&+\tilde{n}_1^4 F_2 (-36 \beta ^2+27 \beta  F_2^2-5 F_2^4) u_{\mathrm{th}}^{10}+48 \tilde{F}_0^4 F_2^2 (F_2
   u_{\mathrm{th}}^2+2)
   \\&+9 \tilde{F}_0 \tilde{n}_1^3 F_2^3 (F_2^2-3 \beta ) u_{\mathrm{th}}^9-4 \tilde{F}_0^2 \tilde{n}_1^2 F_2 (F_2^2-3
   \beta ) u_{\mathrm{th}}^4 (F_2^2 u_{\mathrm{th}}^4+2)],
    \end{aligned}
\end{equation}
where $\mathcal{D}$ is defined as 
\begin{equation}
    \mathcal{D}\equiv\tilde{n}_1 F_2 \left(F_2^2-3 \beta \right) u_{\mathrm{th}}-\tilde{F}_0 F_2^3
\end{equation}

\bibliographystyle{JHEP}
\bibliography{biblio.bib}

\providecommand{\href}[2]{#2}\begingroup\raggedright\begin{thebibliography}{10}

\bibitem{Einstein:1935tc}
A.~Einstein and N.~Rosen, \emph{{The Particle Problem in the General Theory of Relativity}}, \href{https://doi.org/10.1103/PhysRev.48.73}{\emph{Phys. Rev.} {\bfseries 48} (1935) 73}.

\bibitem{Morris:1988cz}
M.S.~Morris and K.S.~Thorne, \emph{{Wormholes in space-time and their use for interstellar travel: A tool for teaching general relativity}}, \href{https://doi.org/10.1119/1.15620}{\emph{Am. J. Phys.} {\bfseries 56} (1988) 395}.

\bibitem{DeFalco:2023twb}
V.~De~Falco and S.~Capozziello, \emph{{Static and spherically symmetric wormholes in metric-affine theories of gravity}}, \href{https://doi.org/10.1103/PhysRevD.108.104030}{\emph{Phys. Rev. D} {\bfseries 108} (2023) 104030} [\href{https://arxiv.org/abs/2308.05440}{{\ttfamily 2308.05440}}].

\bibitem{Harko:2013yb}
T.~Harko, F.S.N.~Lobo, M.K.~Mak and S.V.~Sushkov, \emph{{Modified-gravity wormholes without exotic matter}}, \href{https://doi.org/10.1103/PhysRevD.87.067504}{\emph{Phys. Rev. D} {\bfseries 87} (2013) 067504} [\href{https://arxiv.org/abs/1301.6878}{{\ttfamily 1301.6878}}].

\bibitem{Fierro:2018rna}
O.~Fierro, D.~Narbona, J.~Oliva, C.~Quijada and G.~Rubilar, \emph{{Scalar probes on wormholes in Lovelock theories with unique vacuum}},  \href{https://arxiv.org/abs/1812.02089}{{\ttfamily 1812.02089}}.

\bibitem{Dehghani:2011vu}
M.H.~Dehghani, A.~Bazrafshan, R.B.~Mann, M.R.~Mehdizadeh, M.~Ghanaatian and M.H.~Vahidinia, \emph{{Black Holes in Quartic Quasitopological Gravity}}, \href{https://doi.org/10.1103/PhysRevD.85.104009}{\emph{Phys. Rev. D} {\bfseries 85} (2012) 104009} [\href{https://arxiv.org/abs/1109.4708}{{\ttfamily 1109.4708}}].

\bibitem{Myrzakulov:2015kda}
R.~Myrzakulov, L.~Sebastiani, S.~Vagnozzi and S.~Zerbini, \emph{{Static spherically symmetric solutions in mimetic gravity: rotation curves and wormholes}}, \href{https://doi.org/10.1088/0264-9381/33/12/125005}{\emph{Class. Quant. Grav.} {\bfseries 33} (2016) 125005} [\href{https://arxiv.org/abs/1510.02284}{{\ttfamily 1510.02284}}].

\bibitem{Battista:2024gud}
E.~Battista, S.~Capozziello and A.~Errehymy, \emph{{Generalized uncertainty principle corrections in Rastall-Rainbow Casimir wormholes}},  \href{https://arxiv.org/abs/2409.09750}{{\ttfamily 2409.09750}}.

\bibitem{DeFalco:2021ksd}
V.~De~Falco, E.~Battista, S.~Capozziello and M.~De~Laurentis, \emph{{Reconstructing wormhole solutions in curvature based Extended Theories of Gravity}}, \href{https://doi.org/10.1140/epjc/s10052-021-08958-4}{\emph{Eur. Phys. J. C} {\bfseries 81} (2021) 157} [\href{https://arxiv.org/abs/2102.01123}{{\ttfamily 2102.01123}}].

\bibitem{DiGrezia:2017daq}
E.~Di~Grezia, E.~Battista, M.~Manfredonia and G.~Miele, \emph{{Spin, torsion and violation of null energy condition in traversable wormholes}}, \href{https://doi.org/10.1140/epjp/i2017-11799-6}{\emph{Eur. Phys. J. Plus} {\bfseries 132} (2017) 537} [\href{https://arxiv.org/abs/1707.01508}{{\ttfamily 1707.01508}}].

\bibitem{Ellis:1973yv}
H.G.~Ellis, \emph{{Ether flow through a drainhole - a particle model in general relativity}}, \href{https://doi.org/10.1063/1.1666161}{\emph{J. Math. Phys.} {\bfseries 14} (1973) 104}.

\bibitem{Ellis:1979bh}
H.G.~Ellis, \emph{{The evolving, flowless drain hole: a nongravitating particle model in general relativity theory}}, \href{https://doi.org/10.1007/BF00756794}{\emph{Gen. Rel. Grav.} {\bfseries 10} (1979) 105}.

\bibitem{Bronnikov:1973fh}
K.A.~Bronnikov, \emph{{Scalar-tensor theory and scalar charge}}, {\emph{Acta Phys. Polon. B} {\bfseries 4} (1973) 251}.

\bibitem{Rosa:2022osy}
J.a.L.~Rosa and P.M.~Kull, \emph{{Non-exotic traversable wormhole solutions in linear $f\left( R,T\right) $ gravity}}, \href{https://doi.org/10.1140/epjc/s10052-022-11135-w}{\emph{Eur. Phys. J. C} {\bfseries 82} (2022) 1154} [\href{https://arxiv.org/abs/2209.12701}{{\ttfamily 2209.12701}}].

\bibitem{Rosa:2023guo}
J.a.L.~Rosa, N.~Ganiyeva and F.S.N.~Lobo, \emph{{Non-exotic traversable wormholes in $f\left( R,T_{ab}T^{ab}\right) $ gravity}}, \href{https://doi.org/10.1140/epjc/s10052-023-12232-0}{\emph{Eur. Phys. J. C} {\bfseries 83} (2023) 1040} [\href{https://arxiv.org/abs/2309.08768}{{\ttfamily 2309.08768}}].

\bibitem{Hennigar:2017ego}
R.A.~Hennigar, D.~Kubiz\v{n}\'ak and R.B.~Mann, \emph{{Generalized quasitopological gravity}}, \href{https://doi.org/10.1103/PhysRevD.95.104042}{\emph{Phys. Rev. D} {\bfseries 95} (2017) 104042} [\href{https://arxiv.org/abs/1703.01631}{{\ttfamily 1703.01631}}].

\bibitem{Bueno:2019ycr}
P.~Bueno, P.A.~Cano and R.A.~Hennigar, \emph{{(Generalized) quasi-topological gravities at all orders}}, \href{https://doi.org/10.1088/1361-6382/ab5410}{\emph{Class. Quant. Grav.} {\bfseries 37} (2020) 015002} [\href{https://arxiv.org/abs/1909.07983}{{\ttfamily 1909.07983}}].

\bibitem{Bueno:2022res}
P.~Bueno, P.A.~Cano, R.A.~Hennigar, M.~Lu and J.~Moreno, \emph{{Generalized quasi-topological gravities: the whole shebang}}, \href{https://doi.org/10.1088/1361-6382/aca236}{\emph{Class. Quant. Grav.} {\bfseries 40} (2023) 015004} [\href{https://arxiv.org/abs/2203.05589}{{\ttfamily 2203.05589}}].

\bibitem{Moreno:2023rfl}
J.~Moreno and A.J.~Murcia, \emph{{Classification of generalized quasitopological gravities}}, \href{https://doi.org/10.1103/PhysRevD.108.044016}{\emph{Phys. Rev. D} {\bfseries 108} (2023) 044016} [\href{https://arxiv.org/abs/2304.08510}{{\ttfamily 2304.08510}}].

\bibitem{Hennigar:2016gkm}
R.A.~Hennigar and R.B.~Mann, \emph{{Black holes in Einsteinian cubic gravity}}, \href{https://doi.org/10.1103/PhysRevD.95.064055}{\emph{Phys. Rev. D} {\bfseries 95} (2017) 064055} [\href{https://arxiv.org/abs/1610.06675}{{\ttfamily 1610.06675}}].

\bibitem{Bueno:2016lrh}
P.~Bueno and P.A.~Cano, \emph{{Four-dimensional black holes in Einsteinian cubic gravity}}, \href{https://doi.org/10.1103/PhysRevD.94.124051}{\emph{Phys. Rev. D} {\bfseries 94} (2016) 124051} [\href{https://arxiv.org/abs/1610.08019}{{\ttfamily 1610.08019}}].

\bibitem{Bueno:2016xff}
P.~Bueno and P.A.~Cano, \emph{{Einsteinian cubic gravity}}, \href{https://doi.org/10.1103/PhysRevD.94.104005}{\emph{Phys. Rev. D} {\bfseries 94} (2016) 104005} [\href{https://arxiv.org/abs/1607.06463}{{\ttfamily 1607.06463}}].

\bibitem{Hennigar:2018hza}
R.A.~Hennigar, M.B.J.~Poshteh and R.B.~Mann, \emph{{Shadows, Signals, and Stability in Einsteinian Cubic Gravity}}, \href{https://doi.org/10.1103/PhysRevD.97.064041}{\emph{Phys. Rev. D} {\bfseries 97} (2018) 064041} [\href{https://arxiv.org/abs/1801.03223}{{\ttfamily 1801.03223}}].

\bibitem{Bueno:2017qce}
P.~Bueno and P.A.~Cano, \emph{{Universal black hole stability in four dimensions}}, \href{https://doi.org/10.1103/PhysRevD.96.024034}{\emph{Phys. Rev. D} {\bfseries 96} (2017) 024034} [\href{https://arxiv.org/abs/1704.02967}{{\ttfamily 1704.02967}}].

\bibitem{Hennigar:2017umz}
R.A.~Hennigar, \emph{{Criticality for charged black branes}}, \href{https://doi.org/10.1007/JHEP09(2017)082}{\emph{JHEP} {\bfseries 09} (2017) 082} [\href{https://arxiv.org/abs/1705.07094}{{\ttfamily 1705.07094}}].

\bibitem{Mir:2019rik}
M.~Mir and R.B.~Mann, \emph{{On generalized quasi-topological cubic-quartic gravity: thermodynamics and holography}}, \href{https://doi.org/10.1007/JHEP07(2019)012}{\emph{JHEP} {\bfseries 07} (2019) 012} [\href{https://arxiv.org/abs/1902.10906}{{\ttfamily 1902.10906}}].

\bibitem{Mir:2019ecg}
M.~Mir, R.A.~Hennigar, J.~Ahmed and R.B.~Mann, \emph{{Black hole chemistry and holography in generalized quasi-topological gravity}}, \href{https://doi.org/10.1007/JHEP08(2019)068}{\emph{JHEP} {\bfseries 08} (2019) 068} [\href{https://arxiv.org/abs/1902.02005}{{\ttfamily 1902.02005}}].

\bibitem{Lu:2023hgu}
M.~Lu and R.B.~Mann, \emph{{Maxwell construction and multi-criticality in uncharged generalized quasi-topological black holes}}, \href{https://doi.org/10.1088/1361-6382/ad0db2}{\emph{Class. Quant. Grav.} {\bfseries 41} (2024) 015016} [\href{https://arxiv.org/abs/2306.06733}{{\ttfamily 2306.06733}}].

\bibitem{Lovelock:1970zsf}
D.~Lovelock, \emph{{Divergence-free tensorial concomitants}}, \href{https://doi.org/10.1007/BF01817753}{\emph{Aequat. Math.} {\bfseries 4} (1970) 127}.

\bibitem{Lovelock:1971yv}
D.~Lovelock, \emph{{The Einstein tensor and its generalizations}}, \href{https://doi.org/10.1063/1.1665613}{\emph{J. Math. Phys.} {\bfseries 12} (1971) 498}.

\bibitem{Oliva:2010eb}
J.~Oliva and S.~Ray, \emph{{A new cubic theory of gravity in five dimensions: Black hole, Birkhoff's theorem and C-function}}, \href{https://doi.org/10.1088/0264-9381/27/22/225002}{\emph{Class. Quant. Grav.} {\bfseries 27} (2010) 225002} [\href{https://arxiv.org/abs/1003.4773}{{\ttfamily 1003.4773}}].

\bibitem{Myers:2010ru}
R.C.~Myers and B.~Robinson, \emph{{Black Holes in Quasi-topological Gravity}}, \href{https://doi.org/10.1007/JHEP08(2010)067}{\emph{JHEP} {\bfseries 08} (2010) 067} [\href{https://arxiv.org/abs/1003.5357}{{\ttfamily 1003.5357}}].

\bibitem{Oliva:2012zs}
J.~Oliva and S.~Ray, \emph{{Birkhoff's Theorem in Higher Derivative Theories of Gravity II}}, \href{https://doi.org/10.1103/PhysRevD.86.084014}{\emph{Phys. Rev. D} {\bfseries 86} (2012) 084014} [\href{https://arxiv.org/abs/1201.5601}{{\ttfamily 1201.5601}}].

\bibitem{Oliva:2011xu}
J.~Oliva and S.~Ray, \emph{{Birkhoff's Theorem in Higher Derivative Theories of Gravity}}, \href{https://doi.org/10.1088/0264-9381/28/17/175007}{\emph{Class. Quant. Grav.} {\bfseries 28} (2011) 175007} [\href{https://arxiv.org/abs/1104.1205}{{\ttfamily 1104.1205}}].

\bibitem{Cisterna:2017umf}
A.~Cisterna, L.~Guajardo, M.~Hassaine and J.~Oliva, \emph{{Quintic quasi-topological gravity}}, \href{https://doi.org/10.1007/JHEP04(2017)066}{\emph{JHEP} {\bfseries 04} (2017) 066} [\href{https://arxiv.org/abs/1702.04676}{{\ttfamily 1702.04676}}].

\bibitem{Poisson:2009pwt}
E.~Poisson, \emph{{A Relativist's Toolkit: The Mathematics of Black-Hole Mechanics}}, Cambridge University Press (12, 2009), \href{https://doi.org/10.1017/CBO9780511606601}{10.1017/CBO9780511606601}.

\bibitem{Barriola:1989hx}
M.~Barriola and A.~Vilenkin, \emph{{Gravitational Field of a Global Monopole}}, \href{https://doi.org/10.1103/PhysRevLett.63.341}{\emph{Phys. Rev. Lett.} {\bfseries 63} (1989) 341}.

\bibitem{Lu:2024bdw}
M.~Lu, J.~Yang and R.B.~Mann, \emph{{Gravitational Wormholes}}, \href{https://doi.org/10.3390/universe10060257}{\emph{Universe} {\bfseries 10} (2024) 257} [\href{https://arxiv.org/abs/2406.07734}{{\ttfamily 2406.07734}}].

\bibitem{Radhakrishnan:2024rnm}
R.~Radhakrishnan, P.~Brown, J.~Mutulevich, E.~Davis, D.~Mirfendereski and G.~Cleaver, \emph{{A review of wormhole stabilization in f(R) gravity theories}},  \href{https://arxiv.org/abs/2405.05476}{{\ttfamily 2405.05476}}.

\end{thebibliography}\endgroup

\end{document}